# UNDERSTANDING CRASH DYNAMICS AND SEVERITY OF EV PARATRANSIT: EVIDENCE FROM EASY BIKE ACCIDENTS IN BANGLADESH

**Nazmul Haque[1]***

[1] *Lecturer, Accident Research Institute (ARI),*
*Bangladesh University of Engineering and Technology (BUET), Bangladesh, e-mail: nhaque@ari.buet.ac.bd*

## ABSTRACT

Easy bikes have emerged as a popular and affordable mode of last-mile transport in Bangladesh, yet their widespread use has been accompanied by growing concerns about road safety. This study investigates the underlying factors influencing both the occurrence and severity of easy bike crashes by analyzing nationwide crash data spanning from 2016 to 2024. The findings reveal that crashes predominantly occur during daytime, on paved (pucca) roads, and in low-density peri-urban areas. Intersections and curved road segments are also identified as high-risk zones for crash occurrence. Crash severity analysis, supported by binary logit and probit models, emphasizes that the type of collision plays a crucial role in determining the likelihood of fatal outcomes. Pedestrian-involved crashes and rear-end collisions are more frequently associated with fatalities, whereas crashes involving overturns tend to result in less severe consequences. In contrast, environmental factors such as temperature, rainfall, and time of day exhibit limited impact on crash severity. The distribution of crash types also varies across vehicle categories, with motorcycles, buses, and trucks commonly involved in more dangerous collision scenarios. These findings focus on the urgent need for targeted road safety interventions focusing on specific crash types, high-risk locations such as intersections and curves, and vulnerable road users. Moreover, the study underscores the necessity for improving crash data quality, especially in underreported cases, to support informed decision-making. Based on these insights, the study also proposes a set of evidence-based and context-specific interventions aimed at reducing both the frequency and severity of easy bike crashes in Bangladesh.

***Keywords:*** *Easy Bike, Causation Factors, Crash Severity, Regression, Electric Vehicle (EV), Paratransit*

# 1. INTRODUCTION

The road transport industry is crucial to the movement of both passengers and freight in Bangladesh, a country with a dense population and a rapidly growing economy. The growth of informal and semi-formal transportation options has been a defining characteristic of this environment in recent decades. Among these, the "easy bike," also referred to as an electric three-wheeler or a battery-operated rickshaw, has become a popular and economically significant form of transportation, especially in urban and semi-urban areas (Labib et al. 2013). Millions of people rely on these vehicles for their livelihoods and for affordable mobility, which significantly boosts the local economy and makes many people more accessible (Labib et al. 2013, Singh 2014). An estimated one million to four million battery-powered autorickshaws were in use nationwide by 2024, making them one of the biggest fleets of unofficial electric vehicles in the world (Awal et al. 2019).

Despite their socioeconomic advantages, Easy Bike's explosive and largely uncontrolled growth have been accompanied by a startling rise in traffic fatalities and accidents throughout Bangladesh. With thousands of fatalities each year, road safety continues to be a major public health and development concern. According to reports, three-wheelers—including e-bikes—are a significant contributor to traffic fatalities, accounting for approximately 1.85% of all traffic accident fatalities in 2024. Nevertheless, these accidents have a fatality rate of 1.14 per incident, meaning that more than one person per accident loses their life. This increasing pattern highlights the urgent need for a thorough comprehension of the elements causing these occurrences. Despite this, several studies have examined Easy Bikes' service quality (Paul et al., 2024), energy consumption characteristics (Awal et al., 2019), charging convenience (Uddin et al., 2017), as well as design, construction, and implementation (Salim et al., 2016). However, despite research on Bangladesh's overall road accident rates and contributing factors, there is still a noticeable lack of detailed, targeted studies that focus solely on casualties from e-bikes. By methodically examining the crash occurrence and severity factors associated with easy bike casualties in Bangladesh between 2016 and 2024, this manuscript aims to address this gap. This study aims to identify key trends, common causes, and characteristics of serious outcomes associated with easy bike accidents by analyzing currently available data. The results of this study are essential for guiding regulatory frameworks for e-bikes in Bangladesh, creating targeted road safety campaigns, and informing evidence-based policy interventions. Authorities, legislators, and stakeholders will be able to implement efficient measures to reduce risks, improve vehicle safety standards, enhance driver training, and ultimately decrease the number of fatalities and injuries associated with easy bike accidents on the country's roads if they have a better understanding of these factors.

# 2. METHODOLOGY

## 2.1 Study design

### 2.1.1 Data Sources

Crash records of easy bike accidents were compiled from Bangladeshi newspapers and online news portals published between 2016 and 2024. Data were sourced from eight major outlets: Jugantor, Bangladesh Pratidin, NTV, Prothom Alo, Samakal, Amader Shomoy, Ajker Potrika, and BD24 Live. Among them, Jugantor contributed the highest number of reports (527), suggesting more consistent or comprehensive coverage of traffic incidents. Other significant contributors included NTV (88), Bangladesh Pratidin (48), and Prothom Alo (38), while Amader Shomoy (21), Samakal (20), Ajker Potrika (9), and BD24 Live (4) provided smaller but relevant datasets. These sources were selected for their broad readership, online accessibility, and established record of reporting road traffic events.

The collected crash data are systematically maintained in the specialized crash database of the Accident Research Institute (ARI) at Bangladesh University of Engineering and Technology (BUET). Trained staff members monitor newspapers and online news portals daily to identify and flag easy

bike–related crashes for entry. To ensure accuracy and consistency, three dedicated data entry operators record the information, followed by a rigorous multi-level verification process. Each day's entries are cross-checked by an independent operator who was not involved in the original data entry, helping to identify and correct any discrepancies or omissions. The verifying operator then confirms that all corrections are implemented before final approval. Additionally, a comprehensive validation is performed monthly to ensure the long-term integrity and reliability of the data.

Following collection, crash location data were geocoded using the Google Maps API to obtain latitude–longitude coordinates and convert them into GIS-compatible formats. Environmental data—including monthly averages of temperature and humidity—were retrieved from the Bangladesh Meteorological Department (https://live6.bmd.gov.bd/). Road geometry information, such as curvature, intersection type, hierarchy, and surface condition, was extracted from open-source GIS datasets available on the Humanitarian Data Exchange (HDX) platform (https://data.humdata.org/). Road straightness was quantified as the inverse of the standard deviation of curvature along a segment. Population-related indicators, including population density, were obtained from the Bangladesh Bureau of Statistics (https://bbs.gov.bd/ ). All datasets were integrated within the QGIS environment to conduct spatial and adjacency analyses, ensuring a coherent multi-source analytical framework.

## 2.2 Model specifications

### 2.2.1 Statistical Modelling

Logistic regression is a statistical modeling method used to examine the relationship between a binary dependent variable and one or more independent variables (LaValley 2008). Logistic regression models the probability of a categorical outcome, usually coded as 0 or 1, as opposed to linear regression, which predicts a continuous outcome. It is frequently employed when the response variable indicates whether an event occurred or not, such as whether it was fatal or not. As a linear combination of the predictor variables, the model calculates the log-odds of the event's probability. The logistic regression model can be expressed mathematically as follows:

$$\log\left(\frac{p}{1-p}\right) = \beta_0 + \sum_{i=1}^{n} \beta_i x_i \qquad (1)$$

Where:
$p$ is the probability of the outcome (e.g., choosing metro),
$\beta_0$ is the intercept,
$\beta_i$ (i.e., $\beta_1, \beta_2, \ldots \beta_n$) are the coefficients of the explanatory variables $i$,
The coefficients are typically estimated using maximum likelihood estimation (MLE). A positive coefficient implies that an increase in the predictor increases the log-odds (and thus the probability) of the outcome, while a negative coefficient implies the opposite.

In addition to the logistic model, the probit regression model (Cappellari and Jenkins 2003) is also employed to validate the robustness of estimated relationships. Both models serve similar purposes in predicting binary outcomes; however, they differ in the assumed distribution of the error term. While the logit model assumes a logistic distribution of errors, the probit model assumes a standard normal distribution, leading to slightly different interpretations of the coefficients. The probit model expresses the probability of the event occurring as the cumulative distribution function (CDF) of a standard normal variable. Mathematically, it can be represented as:

$$P(Y = 1 \mid X) = \Phi(\beta_0 + \beta_1 X_1 + \beta_2 X_2 + \cdots + \beta_k X_k)$$

where $\Phi(\cdot)$ denotes the standard normal CDF. The coefficients ($\beta_i$) are estimated using maximum likelihood estimation, similar to logistic regression. Compared to the logit model, the probit model often yields comparable marginal effects but assumes a slightly thinner tail, making it useful when extreme probabilities are less frequent. Employing both models in parallel enhances the reliability of the findings by ensuring that the observed relationships are not an artifact of distributional assumptions.

### 2.2.2 Machine Learning Modelling

Multiple decision trees are combined in the Random Forest (RF) ensemble learning algorithm to enhance predictive accuracy and mitigate overfitting in classification or regression tasks (Breiman, 2001). The bootstrap aggregation (bagging) technique is used to train each tree in the forest on a random subset of the training data, and a random subset of predictor variables is considered for splitting at each node. This randomization improves the model's capacity for generalization by lowering the correlation between individual trees. Averaging across all trees (for regression) or majority voting (for classification) yields the final prediction. Because it can handle complex, high-dimensional, and nonlinear interactions among crash-related variables, including behavioral, environmental, and temporal factors, without making rigid distributional assumptions, RF is especially useful in crash analysis. Additionally, it offers measures of variable importance, which enables researchers to pinpoint the most significant predictors influencing the frequency or severity of crashes. For paratransit modes like EasyBikes, where varied driving conditions and mixed-traffic environments are common, RF's robustness, interpretability, and non-parametric nature make it an effective tool for modeling the complex relationships underlying traffic crash data.

## 3. DATA ANALYSIS AND MODELING

### 3.1 Data Interpretation

The histogram analysis of the easy bike crash data against different contextual variables is shown in Figure 1. The x-axis of the figure shows the categories of the contextual variable. Table 1 gives a brief description of Figure 1. As e-bikes become more prevalent in rural and peri-urban transportation systems, the data show a noticeable increase in e-bike-related collisions, fatalities, and injuries throughout Bangladesh between 2016 and 2024. Easy bikes have become a persistent yet dangerous part of the traffic stream, despite showing a sharp increase in the early years, especially until 2019. The subsequent stabilization at high crash counts provides evidence of this. Union, village, and upazila roads—corridors that often lack appropriate geometric design, segregation, and enforcement—are the primary sites of the majority of crashes.

The influence of operational exposure and travel demand is suggested by temporal patterns that display peak crash frequencies during business days and dry, warm months. Furthermore, the frequency of head-on, rear-end, and pedestrian collisions highlights deficiencies in infrastructure and behavior, such as overloading, poor overtaking judgment, and inadequate pedestrian protection. All of these trends point to structural flaws in the way rural road safety is managed, and demand focused changes in vehicle regulations, driver education, and road design to reduce the growing risks of easy bike operations.

Table 1: Data Analytics (2016–2024) Description of Easy Bike Crashes.

| Crash Trend | Fatality Trends |
|---|---|
| • Easy bike crashes surged from 4 in 2016 to 135 in 2019, then partially stabilized (80–120 annually).<br>• COVID-19 caused a temporary decline in 2020.<br>• Continued high counts indicate easy bikes have become a permanent transport mode with persistent safety challenges. | • Fatalities rose sharply from 4 (2016) to 143 (2019), stabilizing at 120+ per year since 2022.<br>• Despite fewer crashes, fatality persistence reveals systemic safety weaknesses—poor crashworthiness, lack of training, weak enforcement.<br>• Increased interaction with faster/heavier vehicles worsens injury severity. |

| Injury Trends | Crash Typology |
|---|---|
| • Injuries increased from 5 (2016) to 164 (2019), dipped in 2020, then peaked at 222 (2022).<br>• Disproportionate rise in injury count suggests higher crash severity due to overloading, poor road design, lack of safety gear, and mixed traffic exposure. | • Head-on and rear-end collisions dominate—linked to unsafe overtaking and lane incursions.<br>• High rate of pedestrian collisions since 2019 reflects unsafe shared corridors.<br>• Loss of control and overturning indicate mechanical instability and overloading. |
| **Temporal Patterns (Day/Week/Month)** | **Environmental Factors** |
| • Crashes distributed evenly across weekdays but higher on Thursday, Friday, and Sunday due to market and leisure travel.<br>• Monthly peaks in October, January, and September correlate with festivals and harvest seasons.<br>• Rainy months (June–August) show fewer crashes due to lower mobility. | • Crashes peak in dry, warm months (26–32°C) and drop during monsoons.<br>• Inverse correlation with rainfall; positive correlation with temperature.<br>• Crashes highest around 80% humidity (pre-monsoon)—suggesting exposure effects.<br>• Environmental monitoring could help predict high-risk periods. |
| **Spatial Distribution** | **Road Class Vulnerability** |
| • High crash frequencies in Barisal, Mymensingh, and Chuadanga due to weak enforcement and easy bike density.<br>• Lower counts in Dhaka–Gazipur–Narayanganj may reflect underreporting or stricter regulation.<br>• "Unspecified" districts mask true risk zones—highlighting data limitations. | • Highest crash density on National and Regional Highways (~20 & ~17 crashes/km).<br>• Union roads moderately affected (~4 crashes/km); lower tiers show fewer but not necessarily safer.<br>• Crash risk increases on higher-order Pucca roads due to speed and limited safety features. |
| **Geometric Design Factors** | **Population Density Influence** |
| • Crashes concentrated on roads with small curve radii (<100 m).<br>• Higher curvature variability correlates with increased crash risk.<br>• Emphasizes need for speed control, warning signage, and improved horizontal alignment. | • Crashes highest in moderate-density areas (1,000–2,000 people/km²)—peri-urban transition zones.<br>• Lower in dense urban areas (>5,000/km²) due to enforcement and modal alternatives.<br>• Risk peaks where infrastructure is inadequate yet easy bike use is high. |
| **Time of Day Distribution** | **Involvement by Counterparty Vehicle** |
| • ~80% of crashes occur during daytime (business and commuting hours).<br>• Night crashes fewer but often more severe due to visibility and fatigue.<br>• Suggests continuous enforcement, both day and night. | • Most collisions involve motorcycles, buses, and trucks, reflecting mixed traffic dynamics.<br>• "Other vehicle" category indicates prevalence of informal/unclassified vehicles.<br>• High pedestrian involvement underscores unsafe road sharing. |
| **Behavioral and Safety Implication** | |
| • Patterns reveal poor overtaking, braking, and lane discipline.<br>• Highlight urgent need for driver training, vehicle safety standards, pedestrian segregation, and enforcement. | |

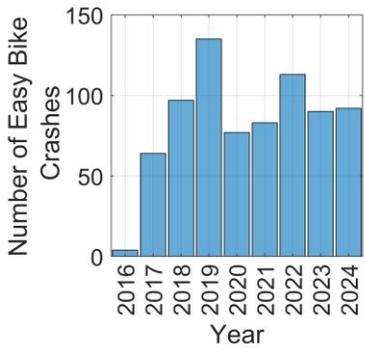
(a)
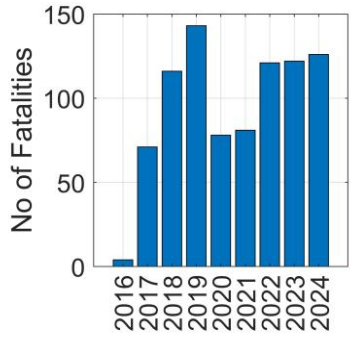
(b)
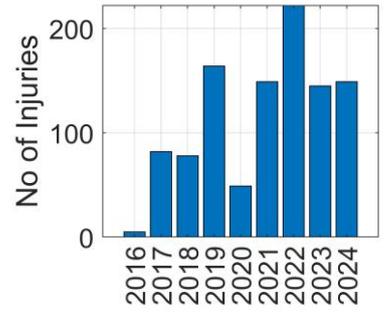
(c)
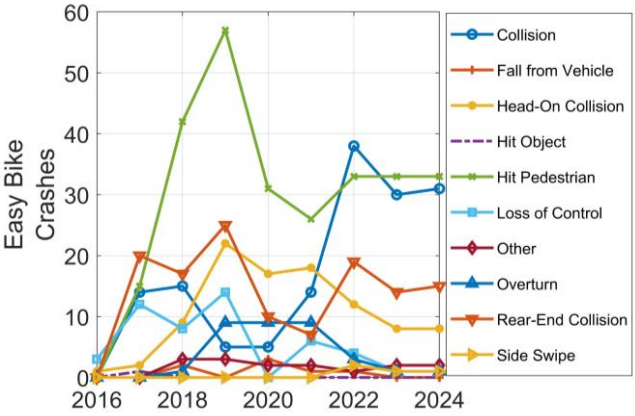
(d)
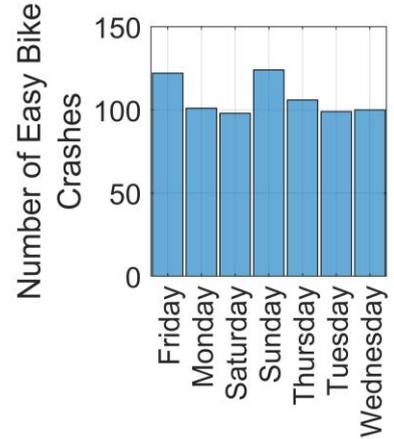
(d)
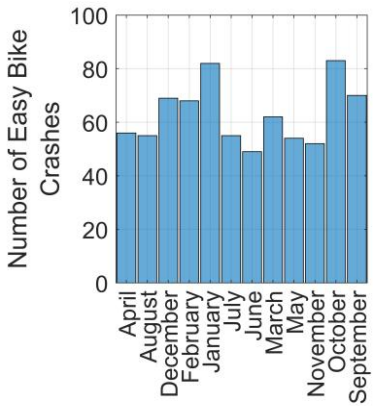
(e)
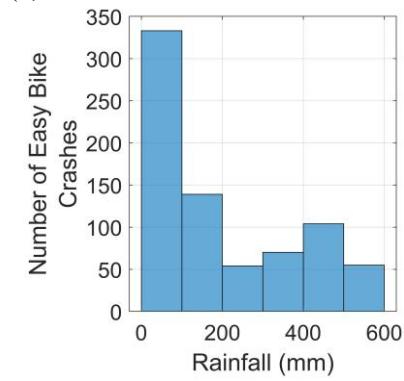
(f)
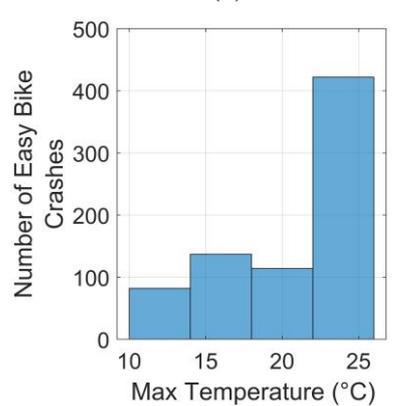
(g)
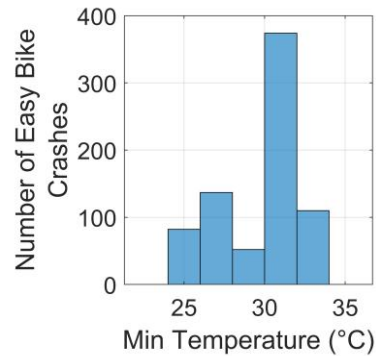
(e)
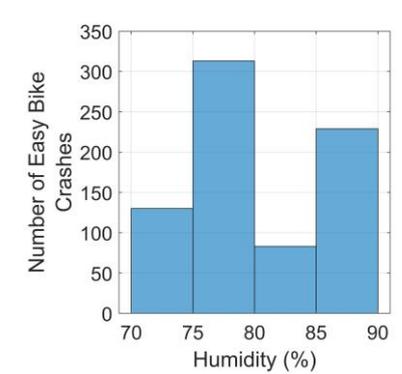
(f)
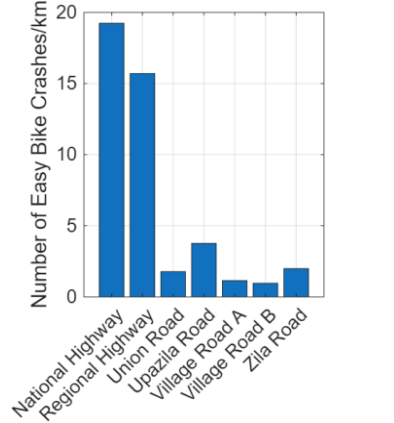
(g)

**6**

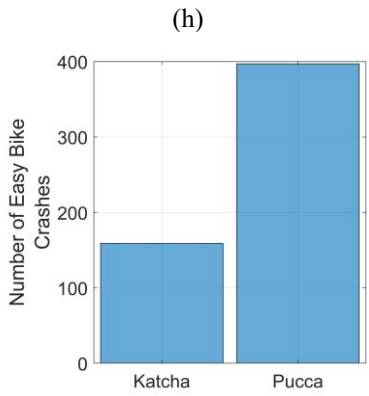
(h)

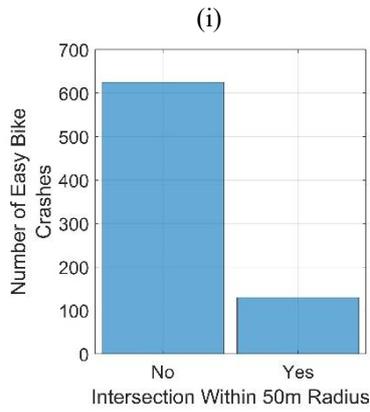
(i)

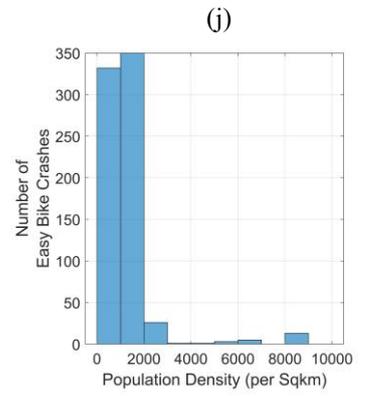
(j)

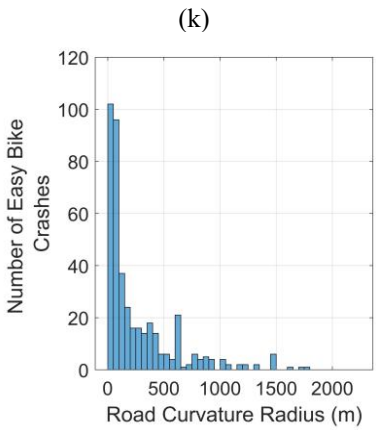
(k)

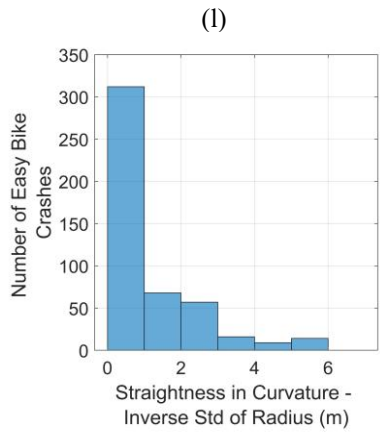
(l)

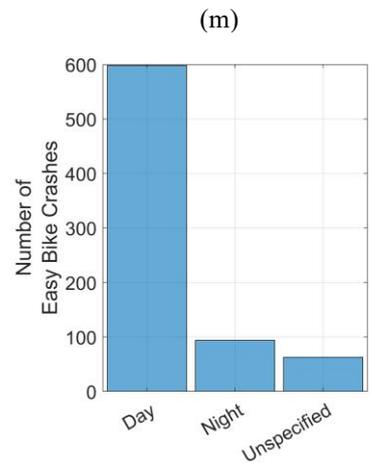
(m)

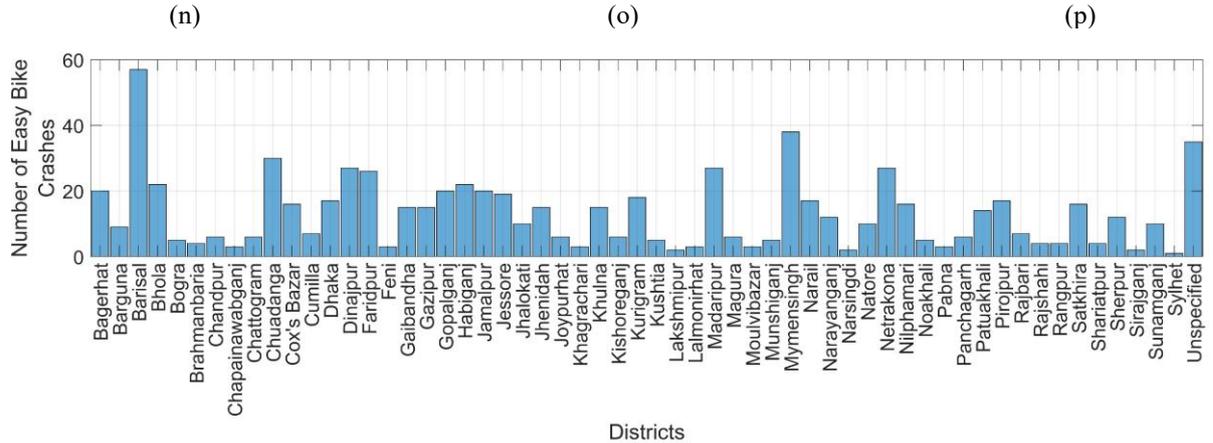
(n)   (o)   (p)

(q)

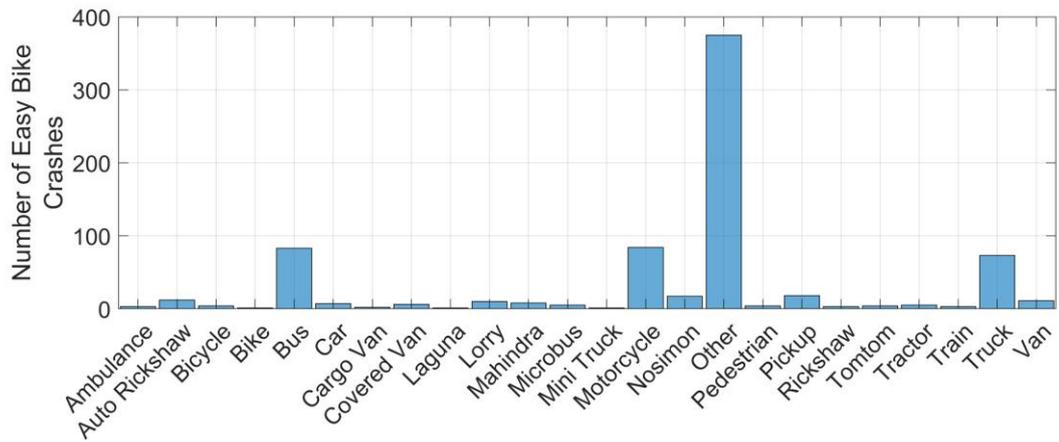

(r)

(s) (t)

Figure 1: Data Analytics (2016–2024) Illustration of the Easy Bike Crashes.

## 3.2 Modelling

Using both binary logistic and probit models, the regression analysis identifies several key factors that influence the fatality outcomes of easy bike crashes in Bangladesh from 2016 to 2024 (Table 2). The analysis reveals that while environmental and geometric conditions appear to have a lesser impact after a crash occurs, temporal factors, crash type, and road characteristics are the primary determinants of crash severity. Temporal variations reveal notable patterns. With logistic and probit coefficients of -0.709 (p = 0.013) and -0.412 (p = 0.009), respectively, crashes on Saturdays are substantially less likely to be fatal. Wednesdays show a similar but weaker trend (logistic = -0.766, p = 0.051), indicating lower odds of fatalities in midweek collisions. These results may indicate slower speeds, reduced exposure to traffic, or different travel habits on weekends compared to weekdays. Furthermore, crashes reported at "unspecified times" have a significant negative correlation with fatal outcomes (logistic = –1.19, p = 0.009), which could be caused in part by reporting gaps because less serious crashes are typically not as thoroughly documented.

November is the safest month according to monthly variation analysis, with logistic and probit coefficients of -1.10 (p = 0.041) and -0.592 (p = 0.034), respectively. This could be because of the favorable weather, more stringent enforcement, or behavioral moderation that occurs during this time. The most important determinant of severity is the type of crash. With logistic and probit coefficients of 3.76 (p < 0.001) and 2.03 (p < 0.001), respectively, pedestrian-involved crashes are substantially more deadly than the reference crash type, with over 40 times the fatality rate. This research highlights the vulnerability of pedestrians in mixed-traffic situations, where slower electric paratransit vehicles, such as e-bikes, collide with faster cars and unprotected drivers. Sparse data and potential problems with perfect prediction limit the statistical reliability of some crash types, such as rear-end, overturn, and stationary-object collisions, despite their exceptionally high coefficients.

The type of road surface also consistently and strongly influences the severity of crashes. With logistic and probit coefficients of 0.781 (p = 0.013) and 0.444 (p = 0.007), crashes on pucca (paved) roads are substantially more deadly than those on katcha (unpaved) roads. According to this pattern, the risk of fatalities is increased by smoother, faster surfaces because they encourage risk-taking and crash energy. Upazila roads have slightly higher fatality odds for crashes, as indicated by road classification (logistic = 0.727, p = 0.077), presumably due to laxer enforcement and subpar design standards. Systemic safety outcomes have not yet changed significantly as a result of the pandemic and legislative factors like the Road Safety Law (RSFLaw) and the COVID-19 period, which have mild but statistically insignificant effects on crash severity. However, it was discovered that neither geometric indicators (road curvature, straightness index) nor environmental factors (temperature, rainfall, and humidity) significantly affected the number of fatalities, indicating that their main impact was on the frequency of crashes rather than the severity of those that followed.

Table 2: Crash Severity Model Results of Easy Bike Crashes

| Coefficient/Variables | Logit | | | | Probit | | | |
|---|---|---|---|---|---|---|---|---|
| | Estimate | SE | tStat | pValue | Estimate | SE | tStat | pValue |
| (Intercept) | 0.000 | 0.000 | - | - | 0.000 | 0.000 | - | - |
| **Day of Week** | | | | | | | | |
| *Friday* | | | | *Reference* | | | | |
| *Monday* | 1.947 | 1.256 | 1.550 | 0.121 | 1.017 | 0.663 | 1.534 | 0.125 |
| *Saturday* | -2.111 | 1.071 | -1.972 | 0.049 | -1.152 | 0.572 | -2.013 | 0.044 |
| *Sunday* | 0.780 | 1.401 | 0.557 | 0.577 | 0.319 | 0.731 | 0.436 | 0.663 |
| *Thursday* | 1.003 | 1.594 | 0.629 | 0.529 | 0.278 | 0.758 | 0.366 | 0.714 |
| *Tuesday* | -0.875 | 1.268 | -0.690 | 0.490 | -0.538 | 0.656 | -0.820 | 0.412 |
| *Wednesday* | -2.270 | 1.243 | -1.826 | 0.068 | -1.272 | 0.660 | -1.928 | 0.054 |
| **Time of Accident** | | | | | | | | |
| *Day* | | | | *Reference* | | | | |
| *Night* | -0.595 | 1.066 | -0.558 | 0.577 | -0.309 | 0.567 | -0.546 | 0.585 |
| *Unspecified* | -2.664 | 1.075 | -2.478 | 0.013 | -1.404 | 0.576 | -2.438 | 0.015 |
| **Month** | | | | | | | | |
| *April* | | | | *Reference* | | | | |
| *August* | 0.498 | 1.379 | 0.361 | 0.718 | 0.108 | 0.763 | 0.141 | 0.888 |
| *December* | -1.439 | 2.043 | -0.705 | 0.481 | -0.929 | 1.073 | -0.866 | 0.387 |
| *February* | 82.667 | * | 0.000 | 1.000 | 13.136 | * | 0.000 | 1.000 |
| *January* | 0.000 | 0.000 | - | - | 0.000 | 0.000 | - | - |
| *July* | 0.000 | 0.000 | - | - | 0.000 | 0.000 | - | - |
| *June* | 1.542 | 1.481 | 1.041 | 0.298 | 0.676 | 0.742 | 0.910 | 0.363 |
| *March* | 1.786 | 2.028 | 0.880 | 0.379 | 0.942 | 1.055 | 0.893 | 0.372 |
| *May* | 0.624 | 1.179 | 0.529 | 0.597 | 0.280 | 0.633 | 0.442 | 0.659 |
| *November* | -3.837 | 1.451 | -2.644 | 0.008 | -2.162 | 0.781 | -2.769 | 0.006 |
| *October* | 0.000 | 0.000 | - | - | 0.000 | 0.000 | - | - |
| *September* | -1.088 | 1.420 | -0.766 | 0.444 | -0.530 | 0.795 | -0.668 | 0.504 |
| **Crash Type** | | | | | | | | |
| *Collision* | | | | *Reference* | | | | |
| *Fall from Vehicle* | 101.877 | * | 0.000 | 1.000 | 16.266 | * | 0.000 | 1.000 |
| *Head-On Collision* | 1.298 | 0.917 | 1.416 | 0.157 | 0.586 | 0.482 | 1.215 | 0.225 |
| *Hit Object* | 96.045 | * | 0.000 | 1.000 | 12.688 | * | 0.000 | 1.000 |
| *Hit Pedestrian* | 4.241 | 1.111 | 3.815 | 0.000 | 2.298 | 0.574 | 4.001 | 0.000 |
| *Loss of Control* | 0.031 | 1.012 | 0.031 | 0.976 | 0.015 | 0.536 | 0.028 | 0.978 |
| *Other* | 1.550 | 1.705 | 0.909 | 0.363 | 0.809 | 0.899 | 0.899 | 0.369 |
| *Overturn* | 0.639 | 1.359 | 0.470 | 0.638 | 0.308 | 0.755 | 0.408 | 0.683 |
| *Rear-End Collision* | 91.537 | * | 0.000 | 1.000 | 15.789 | * | 0.000 | 1.000 |
| *Side Swipe* | 101.842 | * | 0.000 | 1.000 | 15.773 | * | 0.000 | 1.000 |
| **Temperature** | | | | | | | | |
| *Max Temperature* | 0.349 | 0.500 | 0.698 | 0.485 | 0.171 | 0.271 | 0.633 | 0.527 |
| *Min Temperature* | -0.509 | 0.538 | -0.947 | 0.344 | -0.266 | 0.292 | -0.911 | 0.362 |
| **Environment** | | | | | | | | |
| *Normal Rainfall* | -0.012 | 0.008 | -1.384 | 0.166 | -0.006 | 0.005 | -1.364 | 0.173 |
| *Normal humidity* | 0.123 | 0.103 | 1.195 | 0.232 | 0.068 | 0.056 | 1.205 | 0.228 |
| **Special Case** | | | | | | | | |
| *Pandemic (yes)* | 2.062 | 1.474 | 1.399 | 0.162 | 1.170 | 0.756 | 1.547 | 0.122 |
| *RSF Law (yes)* | -2.883 | 2.385 | -1.209 | 0.227 | -1.583 | 1.245 | -1.271 | 0.204 |
| **Road Hierarchy** | | | | | | | | |
| *National Highway* | | | | *Reference* | | | | |
| *Regional Highway* | 81.853 | * | 0.000 | 1.000 | 15.499 | * | 0.000 | 1.000 |
| *Union Road* | 1.186 | 1.327 | 0.894 | 0.371 | 0.742 | 0.696 | 1.067 | 0.286 |
| *Upazila Road* | 3.905 | 2.222 | 1.758 | 0.079 | 2.257 | 1.199 | 1.882 | 0.060 |
| *Village Road A* | 1.889 | 1.342 | 1.408 | 0.159 | 1.187 | 0.713 | 1.665 | 0.096 |
| *Village Road B* | 0.090 | 1.585 | 0.057 | 0.955 | 0.170 | 0.824 | 0.207 | 0.836 |
| *Zila Road* | -0.345 | 1.207 | -0.286 | 0.775 | -0.148 | 0.615 | -0.241 | 0.809 |
| **Road Surface Type (Pucca = true)** | 2.971 | 1.135 | 2.618 | 0.009 | 1.645 | 0.593 | 2.775 | 0.006 |
| **Intersection = true** | 0.398 | 0.919 | 0.433 | 0.665 | 0.224 | 0.507 | 0.441 | 0.659 |
| **Population Density** | 0.000 | 0.000 | 0.318 | 0.751 | 0.000 | 0.000 | 0.496 | 0.620 |
| **Road Geometry** | | | | | | | | |
| *Average Radius* | 0.001 | 0.001 | 0.854 | 0.393 | 0.001 | 0.001 | 0.751 | 0.453 |
| *Straightness* | 0.020 | 0.042 | 0.482 | 0.630 | 0.012 | 0.023 | 0.537 | 0.591 |
| | | | **Goodness of Fit** | | | | | |
| **Ordinary Rsqured** | | | | 0.452101 | | | | 0.41959 |
| **Adjusted Rsqured** | | | | 0.390046 | | | | 0.35385 |
| **AIC** | | | | 184.99 | | | | 185.68 |
| **BIC** | | | | 369.06 | | | | 369.75 |
| **CAIC** | | | | 415.06 | | | | 415.75 |

The random forest modeling is conducted (Precision = 0.22; Recall = 0.4; F1 = 0.29; PCC = 82%) to find the variable influence. Figure 2(a) ranks the predictors based on their contribution to the model's predictive accuracy. The higher the importance value, the more influential the variable is in classifying whether a crash is fatal or not. With the highest relative importance (approximately 0.5), Crash Type stands out as the most significant factor. The nature of the crash (e.g., pedestrian-involved, rear-end, or overturn) is the strongest predictor of fatality risk, according to the regression results. Additionally, road type and straightness (inverse standard deviation of curvature) exhibit high importance values (approximately 0.4–0.45), indicating that the geometry and classification of the roadway have a significant impact on crash results.

Roads that are straighter and have a smoother curvature may encourage faster speeds, which would make crashes more severe. Indicating that regional and geometric features also influence fatality risk,

Population Density and Average Curvature Radius emerge as secondary but significant predictors. Month, the time of the accident, and temperature are examples of temporal and environmental variables that exhibit moderate importance, suggesting that although they affect crash likelihood, they have a relatively smaller direct impact on severity. However, factors such as the pandemic period and road safety law are of very little significance, indicating that changes related to the pandemic or policy interventions have not yet had a discernible impact on crash severity patterns.

Individual predictors have modest effects, but when combined, their collective influence provides clear interpretability, as shown in Figure 2(b). The strongest positive contributor, Crash Type (0.00056), indicates that certain crash types significantly increase the risk of fatalities, particularly rear-end and pedestrian collisions. The fatality risk is also increased by Road Type (0.00089) and Functional Type (0.00047), suggesting that wider or higher-order roads encourage faster operating speeds. On the other hand, geometric factors such as Average Curvature Radius (-0.00096) and Straightness (-0.00107) exhibit marginally negative effects, indicating that driver reaction times are faster on smoother alignments.

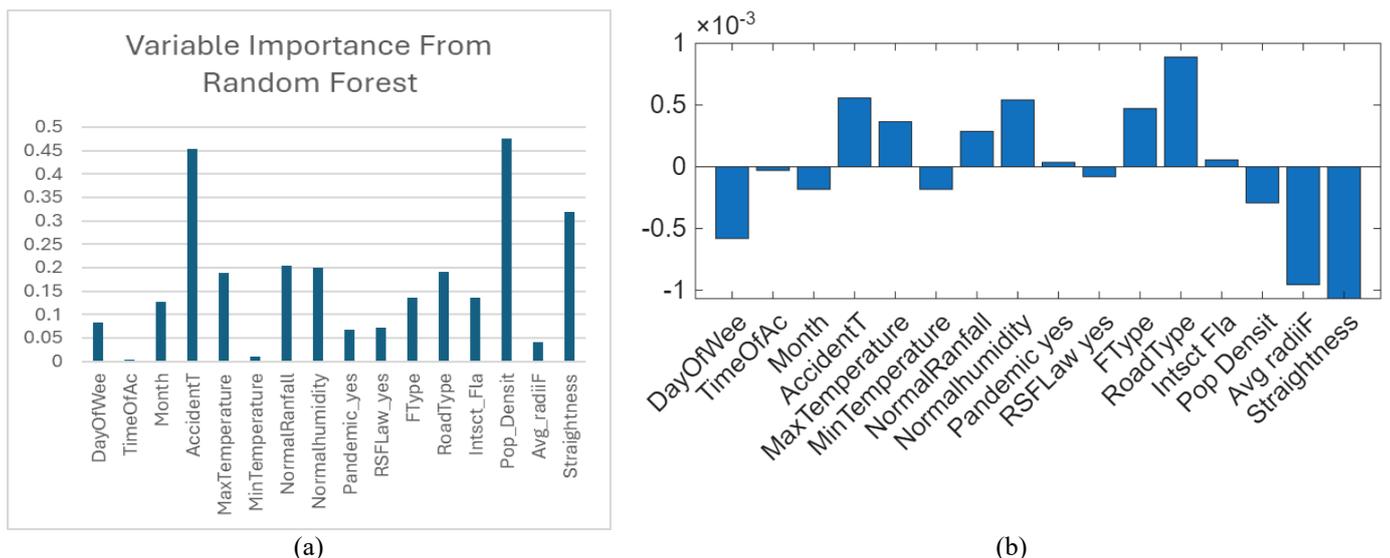

(a) (b)

Figure 2: Variable Influencing the Crash Severity of Easy Bikes.

The risk of fatalities is also reduced by population density (-0.00029), likely due to slower traffic in crowded areas. Temperature, humidity, precipitation, and time-related variables are examples of environmental and temporal variables that show nearly zero SHAP magnitudes, indicating their limited ability to affect crash severity. Overall, the analysis reveals that the most significant factors influencing fatal outcomes in easy bike crashes are the crash mechanism and the roadway context, rather than environmental or temporal variations. This helps policymakers target safety interventions at high-risk crash types and roadway categories.

## 4. CONCLUSIONS AND INTERVENTIONS

This study employed both regression and machine learning techniques to investigate the factors influencing the severity of easy bike crashes in Bangladesh. The results showed that infrastructure and behavior had a significant impact on fatal outcomes. The findings consistently show that the most important predictors of severity are crash type, road geometry, and road surface. While environmental factors like temperature, humidity, and rainfall have little effect, crashes involving pedestrians or those that occur on paved (pucca) and straight road sections are much more likely to be fatal.

These conclusions were supported by the Random Forest and SHAP analyses, which demonstrated that human–infrastructure interactions, rather than climatic or temporal variations, dominate crash fatality risks. Policymakers can benefit significantly from these insights. Segregated lanes or buffer zones for paratransit vehicles, targeted speed management techniques on high-risk paved and straight corridors, and enhanced enforcement and awareness campaigns centered on pedestrian protection could significantly lower the number of fatalities. Additionally, incorporating AI-driven safety monitoring frameworks and fortifying data collection systems would improve proactive decision-making.

Ultimately, this study provides a robust empirical foundation for developing evidence-based road safety regulations, promoting the safer integration of electric paratransit vehicles (EVs) into Bangladesh's mixed traffic environment, and contributing to the nation's broader transition to inclusive and sustainable mobility.

**DECLARATION**

Artificial intelligence (AI) tools were used in the preparation of this manuscript to support language refinement, improve clarity, and assist in structuring the presentation of results. Specifically, *ChatGPT* (OpenAI, GPT-5.1) and *Grammarly* were used to rewrite and refine sections of the text, summarize findings, and enhance narrative coherence. All analytical decisions, interpretations of results, methodological designs, and conclusions were made solely by the authors. The AI tool did not generate or manipulate data, perform statistical analyses, or make any scientific inferences independent of the authors' oversight.